\documentclass[twocolumn]{IEEEtran}
\usepackage{amsmath,amssymb,amsthm,epsfig,color,subfigure,empheq,graphicx,graphics}
\usepackage{enumerate,url,algorithm,algorithmic,wasysym,epstopdf,enumitem,balance}

\newtheorem{lemma}{Lemma}

\newcommand \bzero{\mathbf{0}}
\newcommand \bone{\mathbf{1}}

\newcommand \bd{\mathbf{d}}

\newcommand \bp{\mathbf{p}}
\newcommand \bq{\mathbf{q}}

\newcommand \bs{\mathbf{s}}

\newcommand \bw{\mathbf{w}}
\newcommand \bx{\mathbf{x}}
\newcommand \by{\mathbf{y}}
\newcommand \bz{\mathbf{z}}

\newcommand \bE{\mathbf{E}}
\newcommand \bF{\mathbf{F}}
\newcommand \bG{\mathbf{G}}

\newcommand \bI{\mathbf{I}}

\newcommand \bK{\mathbf{K}}
\newcommand \bL{\mathbf{L}}

\newcommand \bW{\mathbf{W}}

\newcommand \balpha{\boldsymbol{\alpha}}

\newcommand \bomega{\boldsymbol{\omega}}

\newcommand \mcD{\mathcal{D}}
\newcommand \mcE{\mathcal{E}}
\newcommand \mcF{\mathcal{F}}

\newcommand \mcH{\mathcal{H}}

\newcommand \mcP{\mathcal{P}}

\newcommand \hbp{\hat{\mathbf{p}}}

\newcommand \bbd{\bar{\mathbf{d}}}

\newcommand \bbp{\bar{\mathbf{p}}}

\newcommand \bbx{\bar{\mathbf{x}}}

\begin{document}

\title{Data-Driven Modeling of Aggregate Flexibility under Uncertain and Non-Convex Load Models}

\author{
	Sina Taheri,~\IEEEmembership{Member,~IEEE}, 
	Vassilis Kekatos,~\IEEEmembership{Senior Member,~IEEE}, \\
	Harsha Veeramachaneni, and 
	Baosen Zhang,~\IEEEmembership{Member,~IEEE}

    \thanks{S. Taheri and V. Kekatos are with the Bradley Dept. of ECE, Virginia Tech, Blacksburg, VA 24061, USA. S. Veeramachaneni is with NextEra Analytics, Saint Paul, MN 55107, USA. B. Zhang is with the ECE Dept., University of Washington, Seattle, WA 98195. Emails: st92@vt.edu, kekatos@vt.edu, Sriharsha.Veeramachaneni@nexteraanalytics.com,
    and zhangbao@uw.edu. This work was supported by a grant from NextEra Analytics.}
}
\markboth{}{Taheri, Kekatos, Veermachaneni, and Zhang: Data-Driven Modeling of Aggregate Flexibility under Uncertain and Non-Convex Load Models}

\maketitle

\begin{abstract}
Bundling a large number of distributed energy resources through a load aggregator has been advocated as an effective means to integrate such resources into whole-sale energy markets. To ease market clearing, system operators allow aggregators to submit bidding models of simple prespecified polytopic shapes. Aggregators need to carefully design and commit to a polytope that best captures their energy flexibility along a day-ahead scheduling horizon. This work puts forth a model-informed data-based optimal flexibility design for aggregators, which deals with the time-coupled, uncertain, and non-convex models of individual loads. The proposed solution first generates efficiently a labeled dataset of (non)-disaggregatable schedules. The feasible set of the aggregator is then approximated by an ellipsoid upon training a convex quadratic classifier using the labeled dataset. The ellipsoid is subsequently inner approximated by a polytope. Using Farkas’ lemma, the obtained polytope is finally inner approximated by the polytopic shape dictated by the market. Numerical tests show the effectiveness of the proposed flexibility design framework for designing the feasible sets of small- and large-sized aggregators coordinating solar photovoltaics, thermostatically-controlled loads, batteries, and electric vehicles. The tests further demonstrate that it is crucial for the aggregator to consider time-coupling and uncertainties in optimal flexibility design.
\end{abstract}


\section{Introduction}\label{sec:intro}
\allowdisplaybreaks
In addition to the existing bid models for generators and load serving entities, independent system operators (ISO) are currently accepting new bid models to facilitate the participation of load aggregators~\cite{FERC841}. Similar to a battery, new bidding models could consist of upper and lower limits on power and ramping, but also on energy to capture state-of-charge (SoC)-type of constraints. Such battery models capture how \emph{flexible} the aggregator can be as a dispatchable resource for the ISO in the energy market. On one hand, if an aggregator over-estimates its flexibility and then fails to meet its schedule dispatched by the ISO, it will be penalized. On the other hand, offering more flexibility when participating in the market could increase the financial benefit of the aggregator. Evidently, the aggregator aims at offering the maximal flexibility that can be implemented. To achieve this dual goal, the aggregator needs to carefully design its feasibility set to be submitted to the ISO. This task is formally defined as \emph{optimal flexibility design} (OFD).

An aggregator controls a diverse set of devices such as solar photovoltaics (PVs), batteries, electric vehicles (EVs), thermostatically controlled loads (TCLs), home appliances (such as dishwashers/dryers), and pool cleaners/pumps~\cite{SG13}. The OFD task is challenging for three reasons that stem from the properties of such devices or load types. First, devices such as EVs and batteries exhibit naturally time-coupling, meaning that load schedules are constrained across successive control periods and cannot be determined independently. Second, the inefficiencies of EVs and batteries and the ON/OFF characteristics of TCLs lead to non-convex models. Third, most devices operate under time-varying externalities, such as solar irradiance for PVs; the initial SoC for batteries; arrival/departure times for EVs; and ambient temperature or occupancy for TCLs. These externalities are inherently random and uncertain when the OFD problem is solved, e.g., at day-ahead. Such uncertainty further complicates OFD and calls for stochastic formulations.

The current literature in OFD can be categorized in two main groups. The first group of research uses geometric techniques to find a \emph{polytope} that captures the aggregate flexibility~\cite{ASGK14,NTSPV15,HSPV15,Kalsi17,NHBD18,MSSL19,HSAC21}. A reduced-order model for aggregate flexibility is designed using quantization in~\cite{ASGK14}, yet the approach applies only to devices described by convex linear models of the same shape. Stochastic battery models for load aggregations is put forth in~\cite{NTSPV15,HSPV15}, though again they are not applicable to non-convex device models. Presuming externalities to be deterministically known and considering only specific types of devices like batteries and TCLs under averaged linear models, reference~\cite{Kalsi17} models the aggregate flexibility as the Minskowski sum of the convex feasible sets of individual devices, which can be approximated using homothets~\cite{NHBD18}, or zonotopes~\cite{MSSL19}.

The second group of approaches formulates OFD as a multi-level optimization problem~\cite{PM16,SSG18,CL21,CDZL20,CZB21}. Reference~\cite{PM16} models aggregation flexibility by an ellipsoid to be found via semi-definite programming~\cite{PM16}. Reference~\cite{SSG18} considers the cost of flexibility for each device and finds a flexibility cost map, which can be used to find the desired aggregate flexibility region. Nonetheless, the latter approach can handle neither uncertain externalities nor non-convex device models. OFD under linear and deterministic device models has been posed as a multi-stage optimization task also in~\cite{CL21,CDZL20,CZB21}, yet flexibility limits are presumed decoupled across time, which may yield non-implementable aggregator schedules because devices cannot ramp sufficiently fast. Reference~\cite{CZB21} treats disaggregation as a random policy over random externalities, but also ignores time-coupling and non-convex device models. 

In a nutshell, existing works fail to address at least one of the three aforementioned challenges of OFD, namely the time-coupled, non-convex, and stochastic nature of load models. We put forth a data-driven OFD framework that addresses all three challenges. In particular, the contribution of this work is threefold: \emph{1)} Develop a data generation framework that deals with time-coupling in flexibility design, non-convex device models, and uncertain externalities through a chance-constrained formulation (Section~\ref{sec:datagen}); \emph{2)} Train a convex quadratic classifier to approximate the feasible set of the aggregator by an ellipsoid (Section~\ref{sec:learning}); \emph{3)} Inner approximate the obtained ellipsoid with a polytope and use the geometry of polytopes to reformulate OFD as a linear program (Section~\ref{sec:OFD}). Section~\ref{sec:tests} evaluates the performance of the proposed flexibility design solution using two aggregators of increasing modeling complexity. Numerical tests demonstrate that flexibility is well-approximated by convex sets, while capturing time-coupling and uncertainty of load models seems to be important. 

Regarding \emph{notation}, column vectors (matrices) are denoted by lowercase (uppercase) boldface letters; calligraphic symbols are reserved for sets. The $n$-th element of $\bx$ is denoted by $x_{n}$. Symbol $\bone$ denotes the all-one vector. Inequalities between vectors, such as $\bx\geq \by$, apply entry-wise.

\section{Problem Formulation}\label{sec:formulation}
Consider an aggregator participating in a wholesale electricity market cleared by an independent system operator (ISO). To submit its energy bids for the day-ahead market, the aggregator has to comply with the bidding model for virtual generators and flexible loads supported by the ISO. We will henceforth refer to this bidding model as the \emph{aggregator model}. To describe this model, consider a day-ahead market organized in $T$ scheduling intervals indexed by $t$, each one of duration $\delta$. The aggregator model consists of limits $(\underline{p}_t,\overline{p}_t)$ on the instantaneous power $p_t$ provided by the aggregator to the grid during period $t$; its initial state-of-charge (SoC) $s_0$ at the beginning of the first period; limits $\{(\underline{s}_t,\overline{s}_t)\}_{t=1}^{T}$ on its SoC $s_t$ at the end of period $t$; and ramping constraints. The aggregator model is essentially a battery model augmented by ramping constraints. It can be precisely expressed as:
\begin{subequations}\label{eq:battery}
\begin{align}
		&\underline{p}_t\leq p_t\leq\overline{p}_t,\quad && t=1:T\label{eq:battery:p}\\
		&s_{t} = s_{t-1}-\Delta p_t,\quad && t=1:T\label{eq:battery:b}\\
		&\underline{s}_t\leq s_t\leq\overline{s}_t,\quad && t=1:T\label{eq:battery:s}\\
		& \underline{\alpha}_{t}\leq p_{t+1}-p_{t}\leq\overline{\alpha}_{t},\quad && t=1:T-1.
	\end{align}
\end{subequations}
Different from the bidding model of conventional generators, the aggregator model imposes limits on SoC via the additional constraints of~\eqref{eq:battery:s}. Moreover, for conventional thermal generators, capacity and ramping limits are known and typically remain unchanged during normal operations. On the contrary, all limits in \eqref{eq:battery} may be changing across time and day-by-day. The goal of this work is exactly to find these limits.

Let vectors $(\bp,\underline{\bp},\overline{\bp})$ collect the instantaneous power and its limits across all scheduling intervals; vectors $(\underline{\bs},\overline{\bs})$ collect SoC limits; and vectors $(\underline{\balpha},\overline{\balpha})$ collect the ramping rates. The aggregator model is described by \emph{model variables} $(\underline{\bp},\overline{\bp},s_0,\underline{\bs},\overline{\bs},\underline{\balpha},\overline{\balpha})$. To submit a bid to the ISO, the aggregator has to carefully select these variables. Upon collecting bids from all market participants, the ISO clears the market to satisfy demand while ensuring network and reliability constraints. The ISO subsequently informs market participants of their schedules. For the aggregator of interest, the schedule decided by the ISO is denoted by $\bp^*\in\mathbb{R}^T$. This schedule satisfies constraints \eqref{eq:battery} by design. A key objective for the aggregator is to ensure that the scheduled $\bp^*$ can be actually realized, i.e., there exist devices whose dispatches across the scheduling horizon sum up to $\bp^*$. Such schedule will be henceforth termed \emph{disaggregatable} or \emph{feasible}.

In essence, the aggregator model variables define the feasible set wherein the aggregator schedule variable $\bp$ (and hence $\bp^*$) can lie. To express that set in a compact form, eliminate the SoC variables and rewrite \eqref{eq:battery} as the polytope
\begin{equation}\label{eq:P}
\mcP(\bx):=\left\{\bp:\bG\bp\leq\bx\right\}
\end{equation}
where vector $\bx$ depends on the aggregator model variables as
\begin{align}\label{eq:x}
\bx:=\left\{\overline{\bp},-\underline{\bp},\overline{\bs}-s_0\bone,-\underline{\bs}+s_0\bone,\overline{\balpha},\underline{\balpha}\right\}
\end{align}
and the $(6T-2)\times T$ matrix $\bG$ is defined as
\begin{equation}\label{eq:G}
\bG^\top:=[+\bI^\top~-\bI^\top~+\Delta\bL^\top~-\Delta\bL^\top~+\bK^\top~-\bK^\top]
\end{equation}
where $\bI$ is the identity matrix of size $T$; matrix $\bL$ is a $T\times T$ lower triangular matrix with all ones on its lower triangular part; and $\bK$ is a $(T-1)\times T$ difference matrix. Matrix $\bK$ takes the value of $-1$ on its main diagonal; the value of $+1$ on the first above the main diagonal; and zero, otherwise. Because $\bG$ is known and fixed, the ISO only needs to know $\bx$ to dispatch the aggregator. It is therefore fair to say that $\bx$ captures the aggregator flexibility.

While designing $\bx$, the aggregator targets two goals:
\renewcommand{\theenumi}{\emph{\roman{enumi}}}
\begin{enumerate}
    \item Ensure that every $\bp\in\mcP(\bx)$ can be actually implemented by the available resources to avoid penalties; and
    \item Maximize its flexibility while participating in the electricity market.
\end{enumerate}
The latter can be measured by the volume $V(\bx)$ of polytope $\mcP(\bx)$. The task of maximizing flexibility while guaranteeing feasible disaggregation is henceforth termed \emph{optimal flexibility design} (OFD), and can be posed as the optimization
\begin{subequations}\label{eq:maxflex}
	\begin{align}
	\max_{\bx}&~V(\bx)\\
	\mathrm{s.to}&~\mcP(\bx)\subseteq\mcF\label{con:P_in_F}
	\end{align}
\end{subequations}
where $\mcF$ is the set of disaggregatable $\bp$'s. 

\begin{figure*}[t]
	\centering
	\includegraphics[width=0.92\textwidth]{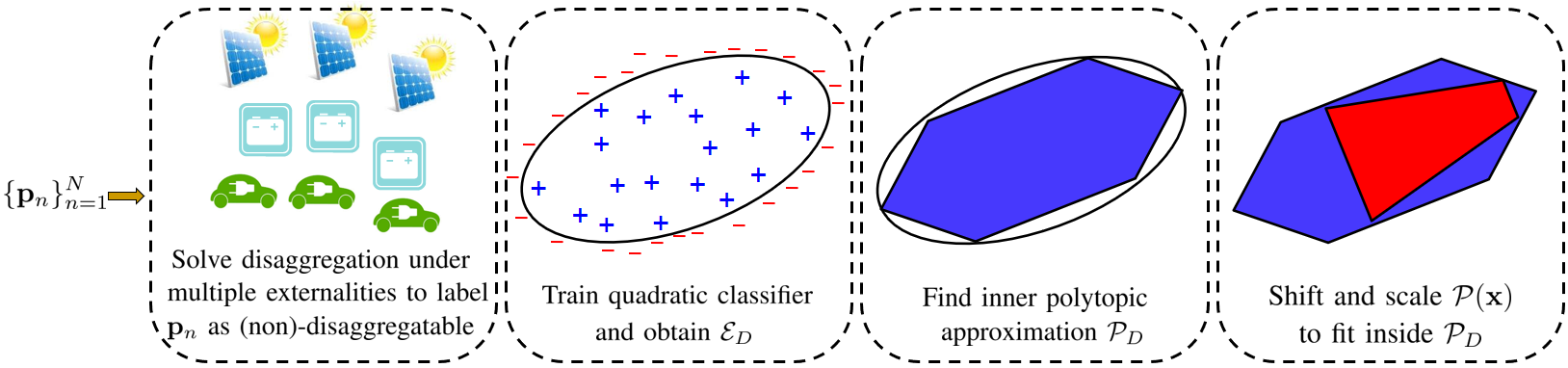}
	\caption{Solving the OFD task in four steps: \emph{S1)} Generating a labeled dataset $\mcD$ of (non)-disaggregatable aggregator schedules by solving disaggregation for multiple instances of externalities $\bomega$; \emph{S2)} Training a convex quadratic classifier to discern disaggregatable schedules and obtaining ellipsoid $\mcE_D$; \emph{S3)} Inner approximating $\mcE_D$ with a polytope $\mcP_D$; \emph{S4)} Finding a maximum volume polytope $\mcP(\bx)$ for a parametric form of $\bx$ to inner approximate $\mcP_D$.}
	\label{fig:flow}
\end{figure*}

Problem~\eqref{eq:maxflex} entails two challenges: First, constraint~\eqref{con:P_in_F} is abstract and it is not obvious how it can handled. Second, finding $V(\bx)$ is hard in general and bears no closed-form mathematical expression. To cope with the first challenge, we take a data-based approach involving the four steps as illustrated in Fig.~\ref{fig:flow}:
\renewcommand{\labelenumi}{\emph{S\arabic{enumi})}}
\begin{enumerate}
    \item Construct a labeled dataset $\mcD:=\{(\bp_n,y_n)\}_{n=1}^N$, where label $y_n=-1$ when $\bp_n$ is disaggregatable; and $y_n=+1$, otherwise (Section~\ref{sec:datagen} and Appendix). 
    \item Use dataset $\mcD$ to approximate the set $\mcF$ of disaggregatable schedules by an ellipsoid $\mcE_D$ (Section~\ref{sec:learning}). 
    \item Find a polytope $\mcP_D$ that inner approximates $\mcE_D$ to arbitrary accuracy (Section~\ref{subsec:poly}).
    \item Design a parametric form for $\bx$ and design it so that $\mcP(\bx)\subseteq\mcP_D$; see Section~\ref{subsec:farkas}. 
\end{enumerate}

Interestingly, step \emph{S4)} introduces a variable that is proportional to the volume of $\mcP(\bx)$. By maximizing this variable, we obviate the need of handling volume $V(\bx)$ explicitly, and thus, address the second challenge. Steps \emph{S1)}--\emph{S4)} are delineated in the next sections. 

\section{Data Generation}\label{sec:datagen}
We commence with step \emph{S1)} of generating the labeled dataset $\mcD$. Note that $\mcD$ may be available to the aggregator from historical data, in which case, step \emph{S1)} is not needed. If historical data are not available or insufficient in numbers, the aggregator can generate training examples by sampling $\bp$'s and labeling them. This section explains how $\bp$'s can be sampled and labeled efficiently. A given $\bp$ is disaggregatable if the disaggregation task presented next yields a zero optimal objective value:
\begin{subequations}\label{eq:DA}
	\begin{align}
	g(\bp;\bomega):=\min_{\{\bp^d\}}~&~\|\bp-\sum_{d=1}^D\bp^d\|\\
	\mathrm{s.to}~&~\bp^d\in\mcF^d(\bomega^d),\quad d=1:D\label{eq:DA:con}
	\end{align}
\end{subequations}
for a vector norm $\|\cdot\|$. Here $D$ is the number of devices controlled by the aggregator. Each vector $\bp^d$ collects the dispatch decisions for device $d$ across all times. The feasible set $\mcF^d(\bomega^d)$ captures the operational constraints $\bp^d$ should satisfy, including for example temperature limits for TCLs, apparent power constraints for PVs, and SoC and power limits for batteries. Each set $\mcF^d(\bomega^d)$ depends on a vector of external parameters $\bomega^d$, such as local ambient temperature for TCLs, solar irradiance for PVs, or arrival/departure times for EVs. Let vector $\bomega$ concatenate all externalities $\bomega^d$'s. The challenges are that \eqref{eq:DA} may not be convex and $\bomega$ is random. 



To arrive at a practical solution, we resort to a sample approximation of the probability of $\bp$ being disaggregatable over uncertain $\bomega$'s. In detail, suppose the aggregator has access to a collection of externality scenarios. For each $\bp$, one randomly samples $K=25$ externality scenarios and solves \eqref{eq:DA} for each oen of them. The sought probability of $\bp$ being disaggregatable can be approximated by the sample statistic
\begin{equation}\label{eq:DA_test}
c(\bp;\Omega_K)=\frac{1}{K}\sum_{k=1}^K\mathbb{I}\left(g(\bp;\bomega_k)=0\right)
\end{equation}
where $\mathbb{I}$ is the indicator function returning one when $g(\bp;\bomega_k)=0$; and zero, otherwise. Schedule $\bp$ is deemed as disaggregatable if $c(\bp;\Omega_K)\geq 1-\epsilon$ for a given small $\epsilon\geq 0$; and infeasible, otherwise. The reason to use only the scenarios in $\Omega_K$ rather than all scenarios in $\Omega_R$ is that solving~\eqref{eq:DA} for $R$ scenarios can be computationally demanding. 

So far we have described a process for determining whether a schedule is disaggregatable or not. In other words, given a $\bp_n$, we can assign it a label $y_n$ by solving $\eqref{eq:DA}$ for $K$ $\bomega$'s and computing the statistic in \eqref{eq:DA_test}. For the classification purposes proposed in the next section, dataset $\mcD$ should be approximately \emph{balanced}, that is disaggregatable (feasible) and non-disaggregatable (infeasible) examples $\bp_n$'s should be similar in numbers. The appendix develops a method that generates efficiently a roughly balanced dataset $\mcD$.

Note by solving \eqref{eq:DA} for a pair $(\bp,\bomega)$, not only we decide whether $g(\bp;\bomega_k)$ is zero; we can also compute the point
\begin{equation}\label{eq:project}
\hbp(\bomega)=\sum_{d=1}^D \hbp^d({\bomega})
\end{equation}
where $\{\hbp^d({\bomega})\}_{d=1}^D$ is the minimizer of \eqref{eq:DA}. Although schedule $\hbp(\bomega)$ can be disaggregated into device schedules for this specific $\bomega$, it is not necessarily disaggregatable in the strict sense used throughout this work as it may not satisfy the chance probability of $c(\hbp(\bomega);\Omega_K)\geq 1-\epsilon$. Schedule $\hbp(\bomega)$ will be useful in the sampling process described in the appendix.

\section{Learning Feasible Aggregations from Data}\label{sec:learning}
This section uses dataset $\mcD$ to approximate $\mcF$ using a convex quadratic classifier. Aggregations of large number of devices with individually non-convex feasible sets can be closely approximated by a convex set~\cite{HSAC21}. However, even if $\mcF$ is non-convex, an ISO would only accept polytopic descriptions for the feasible set of an aggregator to ease its scheduling operations. Therefore, designing a convex classifier as a first step to best capture $\mcF$ is reasonable. Due to the presence of uncertain externalities $\bomega$ and the possibly non-convex nature of $\mcF^d$'s, computing an explicit expression for $\mcF$ is a formidable task. Nonetheless, we can use data to approximate $\mcF$ with a convex set $\hat{\mcF}$. Set $\hat{\mcF}$ can be expressed as the 0-sublevel set of a convex function $d(\bp)$ as $\hat{\mcF}:=\{\bp:d(\bp)\leq 0\}$. Function $d(\bp)$ acts as a \emph{classifier} to decide if $\bp$ is disaggregatable $(d(\bp)\leq0)$ or not $(d(\bp)>0)$.

Prior to learning $d(\bp)$ from $\mcD$, the aggregator needs to choose a functional form for $d(\bp)$. A linear classifier of the form $d(\bp)=\bw_1^\top\bp+w_0$ is the simplest option, but may have limited representation capabilities. A more complex classifier would be a convex quadratic function. Such classifier has been shown to be effective for capturing chance constraints~\cite{LJZ19}, which motivated us to chose the same option as the set $\mcF$ that we are trying to capture here is the feasible set of a chance constraint on $\bp$ as well. While there may be more sophisticated options, such as multi-linear and neural network-based classifiers, they are left for future research. We proceed with training the convex quadratic classifier
\begin{equation}\label{eq:d(p)}
d(\bp):=\bp^\top\bW_2\bp+\bw_1^\top\bp+w_0
\end{equation}
where $\bW_2\succeq 0$ is a symmetric positive semidefinite matrix. The classifier is linear in $(\bW_2,\bw_1,w_0)$ and convex quadratic in $\bp$. It approximates $\mcF$ by the ellipsoid
\begin{equation}\label{eq:ellipsoid}
\mcE_D:=\left\lbrace \bp:\bp^\top\bW_2\bp+\bw_1^\top\bp+w_0\leq 0\right\rbrace.
\end{equation}

The classifier parameters can be learned from dataset $\mcD$ by solving the optimization
\begin{subequations}\label{eq:classifier}
	\begin{align}
	\min~&\frac{1}{N}\sum_{n=1}^N\left[1-y_n d(\bp_n)\right]_{+}+\lambda\left(\|\bW_2\|_F^2+\|\bw_1\|_2^2\right)\label{eq:classifier-cost}\\
	\mathrm{over}~&\bW_2\succeq 0,\bw_1,w_0 \label{con:PSD}
	\end{align}
\end{subequations}
where $[x]_+:=\max(0,x)$ and parameter $\lambda>0$ balances the trade-off between the hinge classification cost under the summation and the regularization terms. Problem~\eqref{eq:classifier} can be reformulated to a semidefinite program (SDP). 

Note that $\mcE_D$ serves only as a surrogate for $\mcF$, and cannot be claimed to be an inner or outer approximation of $\mcF$. This is because $\mcF$ may be non-convex, while $\mcE_D$ has been learned from data and using the sample approximation in \eqref{eq:DA_test}. We next explain how an inner polytopic approximation of $\mcE_D$ can be used to approximate $\mcP(\bx)\subseteq\mcF$ in \eqref{con:P_in_F}.

\section{OFD as Containment of Polytopes}\label{sec:OFD}
Since $\mcE_D$ surrogates set $\mcF$, the constraint $\mcP(\bx)\subseteq\mcF$ in \eqref{con:P_in_F} can be approximated by
\begin{equation}\label{eq:d}
\mcP(\bx)\subseteq\mcE_D\quad\quad \text{or}\quad\quad \max_{\bp:\bG\bp\leq\bx}d(\bp)\leq0.
\end{equation}
Problem~\eqref{eq:d} involves maximizing a convex quadratic function over a polytope. To bypass this non-convexity issue, we proceed in two steps: First find a polytope $\mcP_D$ that is inscribed in $\mcE_D$, and then design $\bx$ so that $\mcP(\bx)\subset\mcP_D\subset \mcE_D$. 

\subsection{Polytopic Inner Approximation of an Ellipsoid}\label{subsec:poly}
We first adopt the approach of \cite{TN01}, \cite{comppoly}, according to which the ellipsoid $\mcE_T^r:=\{\by\in\mathbb{R}^T:\|\by\|_2\leq r\}$ is approximated by a polytope $\mcP_T^\delta$ within accuracy $\delta$ in the sense
\begin{equation*}
\mcE_T^{r/(1+\delta)}\subset\mcP_T^\delta\subset\mcE_T^{r}.
\end{equation*}
Polytope $\mcP_T^\delta$ is defined over the original variables $\by$ and a vector of auxiliary variables $\bq$ as
\begin{equation*}
\mcP_T^\delta:=\left\lbrace \by:\bE_1\by+\bE_2\bq\leq \bbd~~\textrm{for some}~\bq\right\rbrace.
\end{equation*}
The way $(\bE_1,\bE_2,\bbd)$ are determined is summarized in~\cite{SSYG11}. The number of auxiliary variables in $\bq$ and the number of linear constraints in $\mcP_T^\delta$ scale logarithmically with $\delta$. 

Ellipsoid $\mcE_D$ in \eqref{eq:ellipsoid} can be converted to form $\mcE_T^r$ by setting
\[\by:=\bW_2^{1/2}\bp+\bW_2^{-{1/2}}\bw_1~~\textrm{and}~~r:=\bw_1^\top\bW_2^{-1}\bw_1-w_0\]
where $\bW_2^{1/2}$ is the matrix square root of $\bW_2$. Then, ellipsoid $\mcE_D$ can be inner approximated by the polytope
\begin{equation*}
\mcP_D= \left\{\bp:\bE_1(\bW_2^{1/2}\bp+\bW_2^{-1/2}\bw_1)+\bE_2 \bq \leq \bbd\textrm{ for a }\bq\right\}.
\end{equation*}
In words, a point $\bp$ belongs to $\mcP_D$ if there exists a $\bq$ satisfying the aforesaid linear inequalities. In essence, polytope $\mcP_D$ is the projection of a polytope in $(\bp,\bq)$ onto the space of the variables $\bp$ alone. Because such representation of $\mcP_D$ is not convenient for future developments, we next eliminate $\bq$. This projection operation is in general computationally hard given the polytope in $(\bp,\bq)$ is described in its vertex representation. Nevertheless, for moderate lengths of $\bp$ and $\bq$, one can use the Fourier-Motzkin algorithm~\cite{TLIP86}. This algorithm eliminates $\bq$ by generating additional linear constraints on $\bp$. The projection is exact in the sense that one eventually gets the next equivalent representation of $\mcP_D$ for given $(\bE,\bd)$:
\begin{equation}\label{eq:PD}
\mcP_D=\left\lbrace\bp:\bE\bp\leq\bd\right\rbrace.
\end{equation}

\subsection{Reformulating OFD Using Farkas' Lemma}\label{subsec:farkas}
Having found a convenient representation for $\mcP_D$, we can now approximate the OFD problem in \eqref{eq:maxflex} as
\begin{subequations}
\begin{align}\label{eq:maxflex-contain}
\max_{\bx}~&~V(\bx)\\
\mathrm{s.to}~&~\mcP(\bx)\subseteq\mcP_D.\label{eq:maxflex-contain:con}
\end{align}
\end{subequations}
Recall $V(\bx)$ is the volume of $\mcP(\bx)$. We handle constraint \eqref{eq:maxflex-contain:con} upon invoking a version of Farkas' lemma on the containment of polytopes as presented in~\cite{EF82,M02}.

\begin{lemma}\label{lem:Farkas}
[Farkas' lemma]: Consider two non-empty polyhedra $\mcP(\bx)=\left\lbrace\bp:\bG\bp\leq\bx\right\rbrace$ and $\mcP_D:=\left\lbrace\bp:\bE\bp\leq\bd\right\rbrace$. It holds that $\mcP(\bx)\subseteq\mcP_D$ if and only if there exists matrix $\bF\geq\bzero$ satisfying $\bF\bG=\bE$ and $\bF\bx\leq\bd$.
\end{lemma}

Using Lemma~\ref{lem:Farkas}, problem \eqref{eq:maxflex-contain} can be reformulated as
\begin{subequations}\label{eq:maxflex-Farkas1}
	\begin{align}
	\max_{\bx,\bF\geq \bzero}~&~V(\bx)\\
	\mathrm{s.to}~&~\bF\bG = \bE\label{eq:maxflex-Farkas1:con1}\\
	&~\bF\bx\leq\bd.\label{eq:maxflex-Farkas1:con2}
	\end{align}
\end{subequations}
Due to the product $\bF\bx$, problem~\eqref{eq:maxflex-Farkas1} remains non-convex, while we still lack a good choice for $V(\bx)$. 

To resolve these two issues, we resort to a restriction of \eqref{eq:maxflex-Farkas1}. We parameterize the sought vector $\bx$ as
\begin{equation}\label{eq:x_param}
\bx=\frac{1}{\beta}\left(\bbx-\bG\bz\right)
\end{equation}
where $\bz$ and $\beta>0$ are to be designed, whereas $\bbx$ is given. Thanks to this form, constraint \eqref{eq:maxflex-Farkas1:con2} can be written as
\[\frac{1}{\beta}\bF\left(\bbx-\bG\bz\right)\leq \bd ~~\Leftrightarrow~~\bF\bbx\leq \bF\bG\bz+\beta\bd=\bE\bz+\beta\bd\]
using the key observation that $\bF\bG=\bE$ from \eqref{eq:maxflex-Farkas1:con1}. 

\begin{figure}
    \centering
    \includegraphics[scale=0.4]{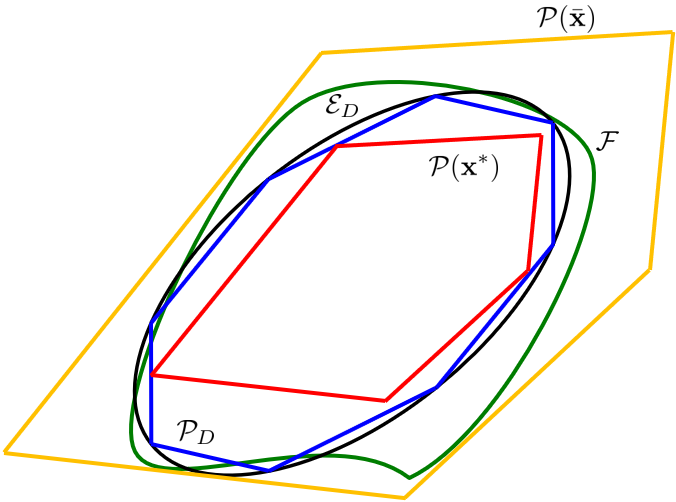}
    \caption{The non-convex original set $\mcF$; the ellipsoid $\mcE_D$ being a data-based estimate of $\mcF$; set $\mcP_D$ is the inner polytopic approximation of $\mcE_D$; the prototype polytope $\mcP(\bar{\bx})$; and the solution polytope $\mcP(\bx^*)$. Notice that $\mcE_D$ and $\mcP(\bar{\bx})$ need not be contained in $\mcF$.}
    \label{fig:feas_set}
\end{figure}

To better understand the proposed restriction, consider polytope $\mcP(\bbx):=\{\bp:\bG\bp\leq\bbx\}$. Evidently under the restriction of \eqref{eq:x_param}, if $\bbp\in\mcP(\bbx)$, then 
$\bp:=(\bbp-\bz)/\beta\in\mcP(\bx)$. This means that if we take any point in $\mcP(\bbx)$, shift it by $-\bz$, and scale it by $1/\beta$, we obtain $\mcP(\bx)$. In other words, the polytope $\mcP(\bx)$ is selected to be a shifted and scaled version of the `prototype' polytope $\mcP(\bbx)$; see also Fig.~\ref{fig:feas_set}. The shape of the prototype polytope depends on $\bbx$, while its shifting and scaling parameters can be optimally selected. 

Adopting the restriction of \eqref{eq:x_param} simplifies the task of dealing with volume $V(\bx)$. Because $\mcP(\bx)\subset \mathbb{R}^T$ is a shifted and scaled replica of $\mcP(\bbx)\subset \mathbb{R}^T$, it holds that
\begin{equation}\label{eq:V(x)}
    V(\bx)=V(\bbx)/\beta^T
\end{equation}
where $V(\bbx)$ is the fixed volume of $\mcP(\bbx)$. Consequently, under \eqref{eq:x_param}, problem~\eqref{eq:maxflex-Farkas1} simplifies to the linear program
\begin{subequations}\label{eq:maxflex-Farkas2}
    \begin{align}
	(\bF^*,\bz^*,\beta^*)\in\arg\min_{\bF\geq \bzero,\bz,\beta\geq 0}~&~\beta\\
	\mathrm{s.to}~&~\bF\bG = \bE\\
	&~\bF\bbx\leq \bE\bz+\beta\bd.
	\end{align}
\end{subequations}

The final answer to the OFD problem is the aggregator model $\bG\bp\leq \bx^*$ with
\begin{equation}\label{eq:x_final}
\bx^*=\frac{1}{\beta^*}\left(\bbx-\bG\bz^*\right)
\end{equation}

Finally, for the parameterization in \eqref{eq:x_param}, we need a suitable choice for $\bbx$. One heuristic would be to find $\bbx$ as the minimum-norm $\bx$ for which $\mcP(\bbx)$ contains all disaggregatable schedules in dataset $\mcD$ by solving
\begin{subequations}\label{eq:xo_minimal}
    \begin{align}
	\bbx=\arg\min_{\bx}~&~\|\bx\|_2^2\\
	\mathrm{s.to}~&~\bG\bp_n\leq\bx,~~\forall~\bp_n\in\mcD~\text{with}~y_n=-1.
	\end{align}
\end{subequations}
Heed that $\mcP(\bbx)$ for the aforesaid choice of $\bbx$ does contain all feasible points in $\mcD$, but it may not lie inside $\mcP_D$ and may contain infeasible points too as illustrated in Fig.~\ref{fig:feas_set}. Table~\ref{alg:OFD} summarizes the steps of the proposed OFD approach.

\begin{algorithm}[t]
	\caption{Optimal Flexibility Design (OFD)}\label{alg:OFD}
	\begin{algorithmic}[1]
		\renewcommand{\algorithmicrequire}{\textbf{Input:}}
		\renewcommand{\algorithmicensure}{\textbf{Output:}} 
		\REQUIRE Dataset of (dis)-aggregatable schedules  $\mcD$.
		\ENSURE Maximum-volume polytope
		$\mcP(\bx^*)$.
		\STATE Train quadratic classifier $d(\bp)$ to find ellipsoid $\mcE_D$ by solving \eqref{eq:classifier}.
		\STATE Find polytope $\mcP_D\subset\mcE_D$ in the $(\bp,\bq)$ representation as described in~\cite{SSYG11}.
		\STATE Convert $\mcP_D$ in the form of \eqref{eq:PD} via the Fourier-Motzkin algorithm.
		\STATE Compute $\bbx$ by solving~\eqref{eq:xo_minimal}.
		\STATE Solve~\eqref{eq:maxflex-Farkas2} and compute the optimal $\bx^*$ from~\eqref{eq:x_param}.
	\end{algorithmic}
\end{algorithm}

\section{Numerical Tests}\label{sec:tests}

\subsection{Simple Illustrative Example}\label{sec:2_time_example}
\begin{figure*}[t]
	\centering
	\subfigure{\includegraphics[width=0.24\textwidth]{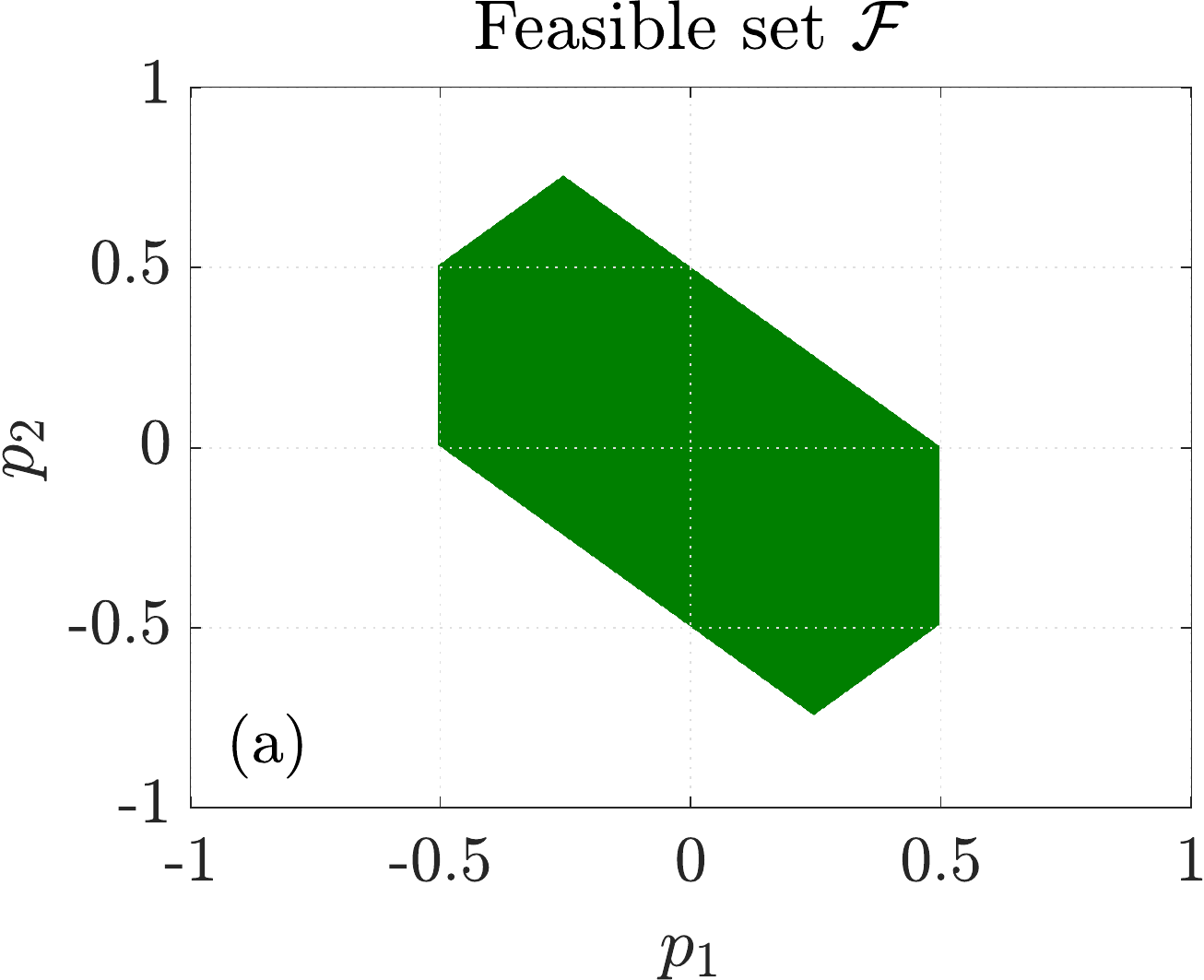}}
	\subfigure{\includegraphics[width=0.24\textwidth]{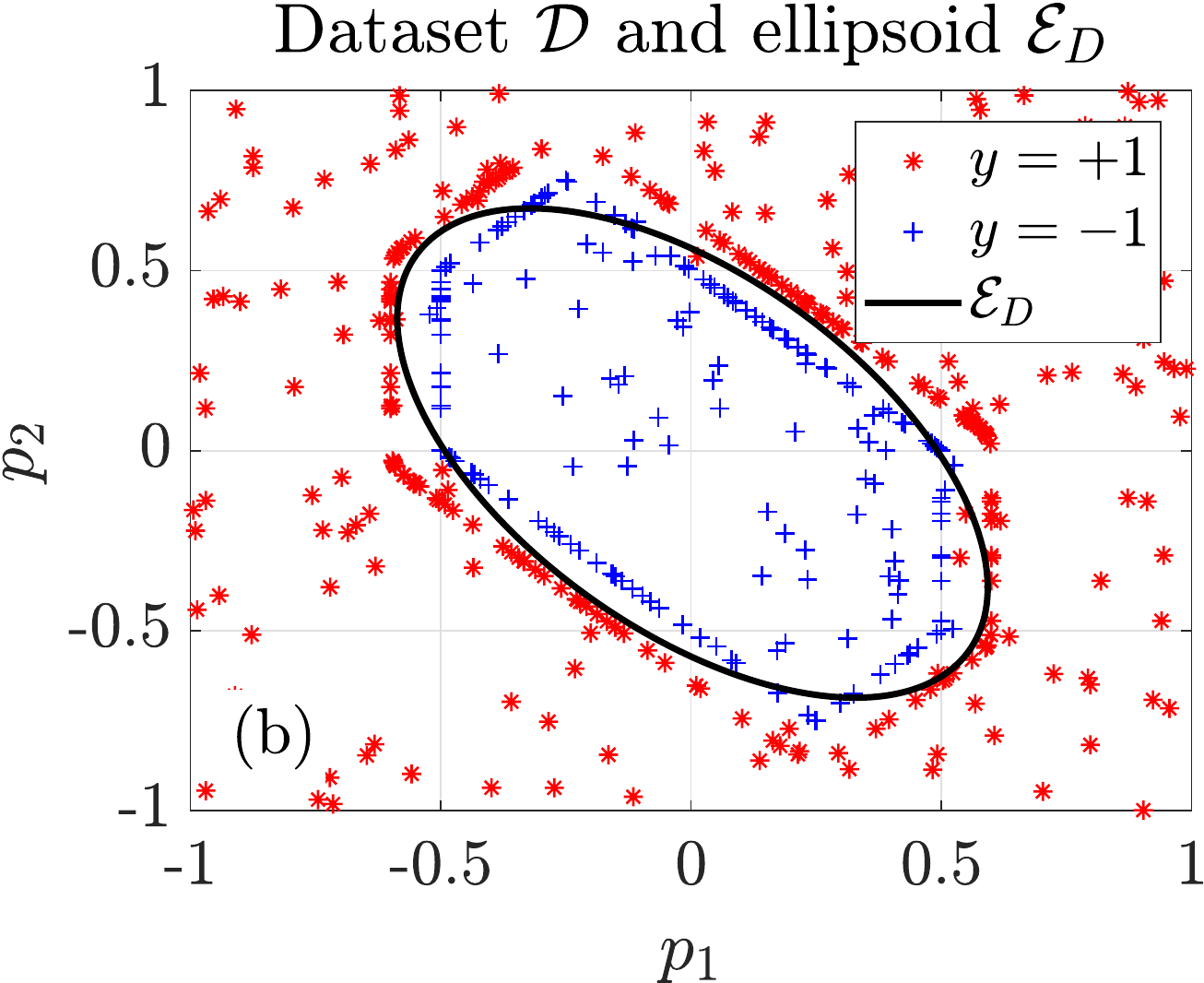}}
	\subfigure{\includegraphics[width=0.24\textwidth]{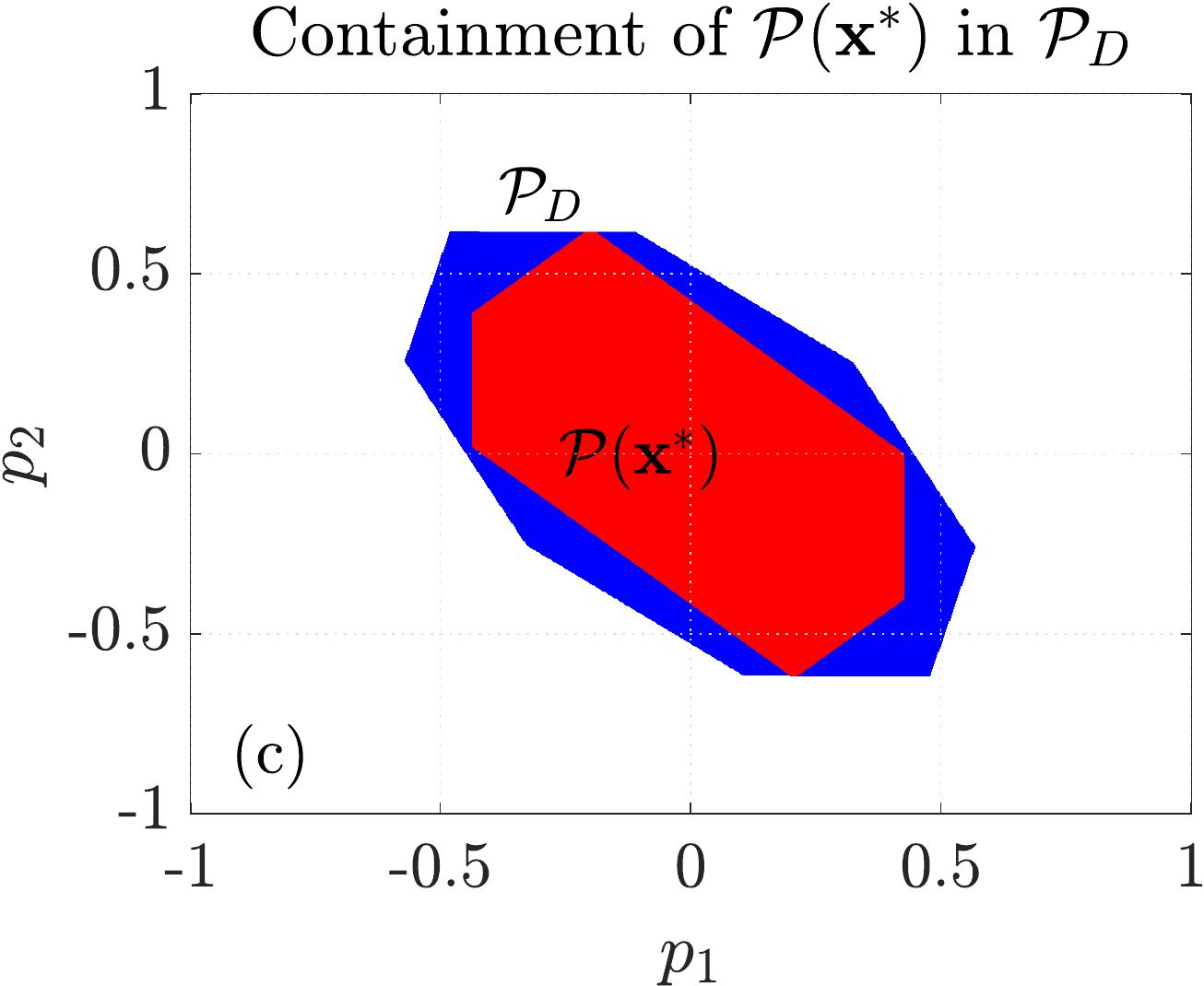}}
	\subfigure{\includegraphics[width=0.24\textwidth]{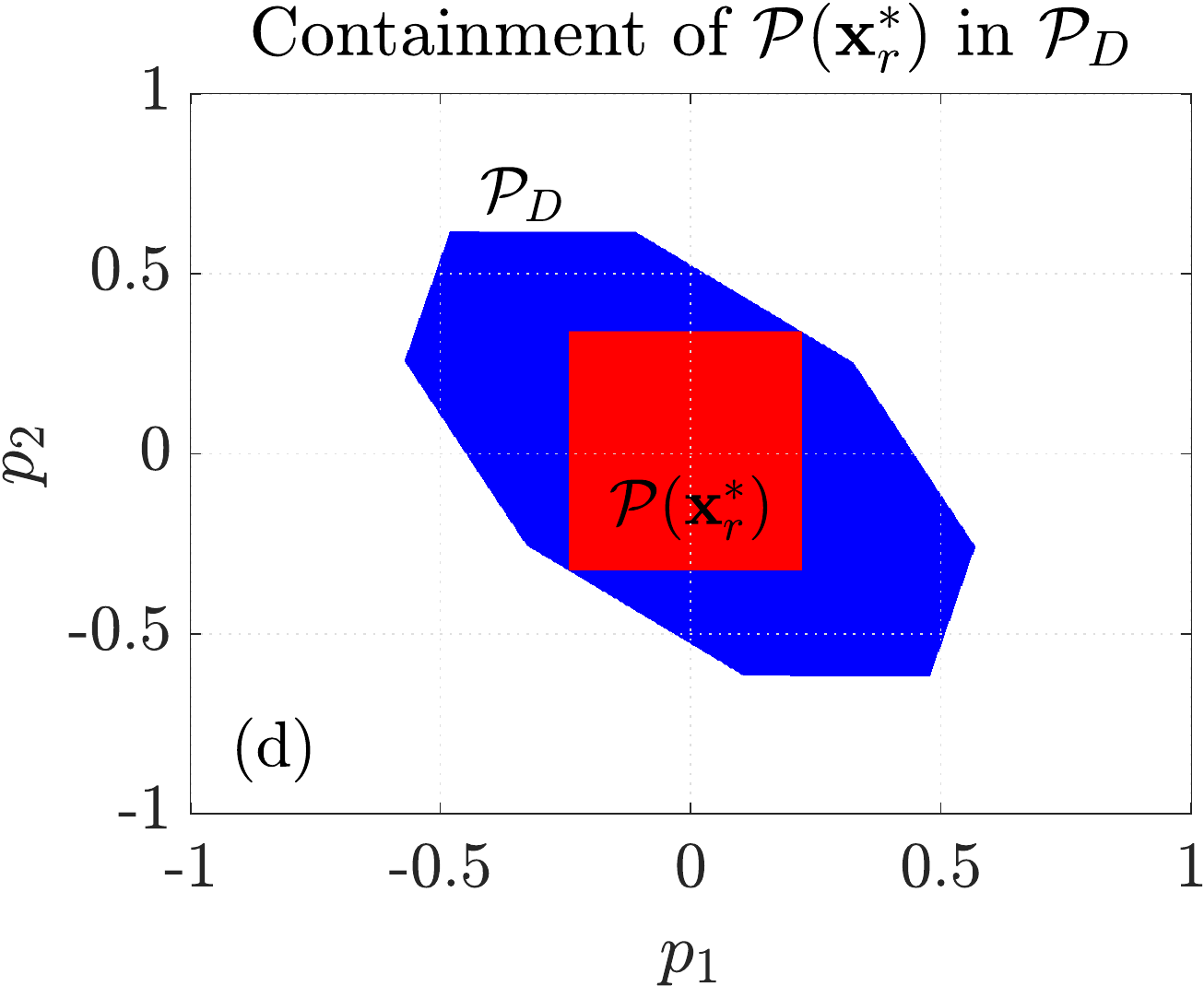}}
	\caption{\emph{(a)} The feasible set $\mcF$ of an aggregator matches with the model $\bG\bp\leq \bx$ of \eqref{eq:P} postulated by the ISO. \emph{(b)} We sampled $N=500$ labeled datapoints (schedules) to train a convex quadratic classifier and obtained ellipsoid $\mcE_D$. \emph{(c)} Inner polytopic approximation $\mcP_D$ of $\mcE_D$, and the final solution $\mcP(\bx^*)$. \emph{(d)} Optimal solution $\mcP(\bx_r^*)$ if the ISO simplifies the model $\bG\bp\leq \bx$ to account only for box constraints and ignore any time coupling. Notice that $\mcP(\bx^*)$ has a much larger volume than $\mcP(\bx_r^*)$, which indicates the importance of more elaborate aggregator models in OFD.}
	\label{fig:T=2}
\end{figure*}

The proposed methodology was first evaluated under a synthetic setup, which has been oversimplified for the purposes of visualization and drawing intuition. This setup involves only $T=2$ control periods, while externalities are deterministic. Moreover, the actual feasible set $\mcF$ of disaggregatable schedules is convex. It in fact matches the aggregator model $\mcP(\bx):=\{\bp:\bG\bp\leq \bx\}$ postulated by the ISO in \eqref{eq:P}. The actual $\bx$ can be computed from \eqref{eq:x} by setting 
\[\overline{\bp}=-\underline{\bp}=\overline{\bs}=\bone_2,~ \underline{\bs}=\bzero_2,~s_0=0.5,~\textrm{and}~\overline{\alpha}=-\underline{\alpha}=1.\]
The feasible set $\mcF$ is shown in Fig.~\ref{fig:T=2}a. Note that out of the $6T-2=10$ linear inequalities in $\mcP(\bx)$, only $6$ are binding for the particular $\bx$. Of course, the aggregator does not know that $\mcF$ is actually $\mcP(\bx)$. It can only draw points $\bp\in\mathbb{R}^2$ and test whether they are disaggregatable. The goal for the aggregator is to fit $\mcP(\bx)$ onto data by finding $\bx$ to maximize its flexibility. 

We first followed the process described in Section~\ref{sec:datagen} and the Appendix to generate a dataset $\mcD$ of $N=500$ points shown in Fig.~\ref{fig:T=2}b. We set parameter $\kappa=0.2$. Observe there are pairs of (in-)feasible points lying close to the boundary of $\mcF$; infeasible points away from the boundary of $\mcF$; and feasible points on the interior of $\mcF$. We then trained a classifier per \eqref{eq:classifier} with $\lambda=10^{-5}$. The obtained ellipsoid $\mcE_D$ is also shown on Fig.~\ref{fig:T=2}b. Ellipsoid $\mcE_D$ closely approximates $\mcF$ as it achieves $99.68\%$ accuracy on the training data. We next followed the steps of Section~\ref{subsec:poly} with $\delta=0.1$ and first obtained the polytope $\mcP_D\subset\mcE_D$ with $\bq\in\mathbb{R}^2$ and then used the Fourier-Motzkin algorithm to convert $\mcP_D$ to the form of~\eqref{eq:PD}. We subsequently ran Algorithm~\ref{alg:OFD} and obtained the solution polytope $\mcP(\bx^*)$. The containment of $\mcP(\bx^*)$ in $\mcP_D$ is depicted in Fig.~\ref{fig:T=2}c. This figure shows that the recovered polytope $\mcP(\bx^*)$ is very close to the original feasible set $\mcF$.

Previous works considered only the lower/upper power limits in \eqref{eq:battery:p}~\cite{CLG18}. By ignoring ramping and SoC limits, they have postulated a simpler aggregator model. To demonstrate the effect of simpler aggregator models, we solved~\eqref{eq:maxflex-Farkas2} again but this time with tried to find a polytope defined by the simpler $\bG^\top=\left[\bI^\top -\bI^\top\right]$ yielding a hyper-rectangle. Figure~\ref{fig:T=2}d illustrates the results for this test. The volume of $\mcP(\bx_r^*)$ is much smaller than the volume of the polytope $\mcP(\bx^*)$ depicted in Fig.~\ref{fig:T=2}c. This demonstrates the importance of considering time-coupling and its effect on the total volume recovered.

\subsection{A More Realistic Aggregator Model}\label{sec:200_device}
We also tested a more realistic setting of an aggregator controlling $200$ loads described by non-convex models subject to random externalities. This aggregator controls PVs, batteries, EVs, and TCLs, $50$ of each type. Although the market is ran on an hourly basis indexed by $t=1,\ldots,T$, loads are controlled at a finer timescale of $15$-min intervals indexed by $\tau=1,\ldots,4T$. This is practical for devices such as TCLs and PVs. If $\ell_\tau^d$ denotes the kW load consumed by device $d$ during the $15$-min interval $\tau$, the kW load for this device over hour $t$ is
\[p_t^d=\frac{1}{4}\sum_{k=1}^4\ell_{4(t-1)+k}^d.\]
We next describe the models loads $\ell_\tau^d$ should satisfy for each one of the four types of loads. These models determine the device feasible sets $\mcF^d(\bomega^d)$ in \eqref{eq:DA}. To keep the notation uncluttered, we will drop superscript $d$. 

\emph{Photovoltaics} were modeled as negative loads constrained as 
\[-r_\tau\bar{p}\leq \ell_\tau \leq 0,\quad\quad \tau=1:4T\]
where $\bar{p}$ is the PV capacity and $r_\tau$ the expected solar irradiance at interval $\tau$. PV capacities were randomly sampled from real data from Southern California Edison (SCE)~\cite{BD19}. To generate profiles for $r_\tau$, we took $1$-min solar irradiance data from \cite{SMART}, perturbed them by $\pm10\%$ with a normal distribution, and averaged them over $15$-min intervals.

\emph{Batteries}: We used a model with charging/discharging inefficiencies described below for $\tau=1\ldots,4T$:
\begin{align*}
	&\ell_\tau= \ell_\tau^{+}-\ell_\tau^{-},\quad 0\leq \ell_\tau^{+}\leq b_\tau\bar{p},\quad
    0\leq \ell_\tau^{-}\leq(1-b_\tau)\bar{p}\\
	&s_\tau=s_{0}+\frac{1}{4} \sum_{k=1}^{\tau}\left(0.9\ell^{+}_k-1.1\ell^{-}_k\right), \quad 0\leq s_\tau\leq\bar{s}
\end{align*}
where $(\ell^{+}_\tau,\ell^{-}_\tau)$ are the (dis)-charging powers during interval $\tau$, and binary variable $b_\tau$ indicates whether this battery is charging or discharging. The binary variable is needed to properly capture inefficiencies. Quantities $(\bar{p},\bar{s})$ are the limits on power and SoC obtained also from SCE data~\cite{BD19}. The initial SoC $s_0$ was drawn independently and uniformly at random within $[0,\bar{s}]$. The $1/4$ term in front of the sum accounts for the $15$-min intervals. 

\emph{Electric vehicles} were modeled similarly to batteries with the additional constraints that $(\ell_\tau,\ell_\tau^+,\ell_\tau^-)$ are zero for intervals $\tau$ the EV is not available. The assumption here is that EVs come and park at business locations. Upon arrival, each EV specifies a departure time and the needed SoC upon departure. Based on the most common EV models in the US market, we chose kW capacities from the set $\bar{\bp}\in \{11, 16.5, 18, 19.2, 20, 21.1, 22\}$~kW, and kWh capacities from the set $\bar{s}\in\{42, 60, 70, 75, 85, 90, 100\}$~kWh. Arrival and departure times were generated by sampling independently from a truncated Gaussian distribution between 9-10 AM and 4-5 PM, respectively. The requested SoC upon departure was drawn uniformly between the SoC upon arrival and the SoC capacity of each vehicle. 

\emph{Thermostatically controlled loads} obeyed the discrete-time model assuming operation under the cooling mode~\cite{Kalsi17}:
\begin{align*}
&0\leq \ell_\tau\leq b_\tau \bar{p},\quad\quad\quad \theta_s-0.5\leq\theta_{\tau}\leq \theta_s+0.5  \\
&\theta_{\tau+1} = \theta_\tau-\frac{1}{4C}\sum_{l=1}^{\tau}\left(P \ell_l-\frac{\theta^a_l}{R}\right).
\end{align*}
Here binary variable $b_\tau$ indicates whether the TCL is on, $\bar{p}$ is the fixed power consumed when the TCL is on, while $\theta_\tau$ and $\theta^a_\tau$ are respectively the house and predicted ambient temperatures in Celcius at time $\tau$. Parameters $(C,P,R)$ are the thermal capacitance, resistance, and power transfer of the house. They were drawn uniformly at random from $[1.5,2.5]$, $[3,5]$, and $[15,30]$, respectively. The temperature setpoint $\theta_s$ was drawn uniformly at random from $[24,26]$. 

\begin{figure}[t]
	\centering
	\subfigure{\includegraphics[width=0.24\textwidth]{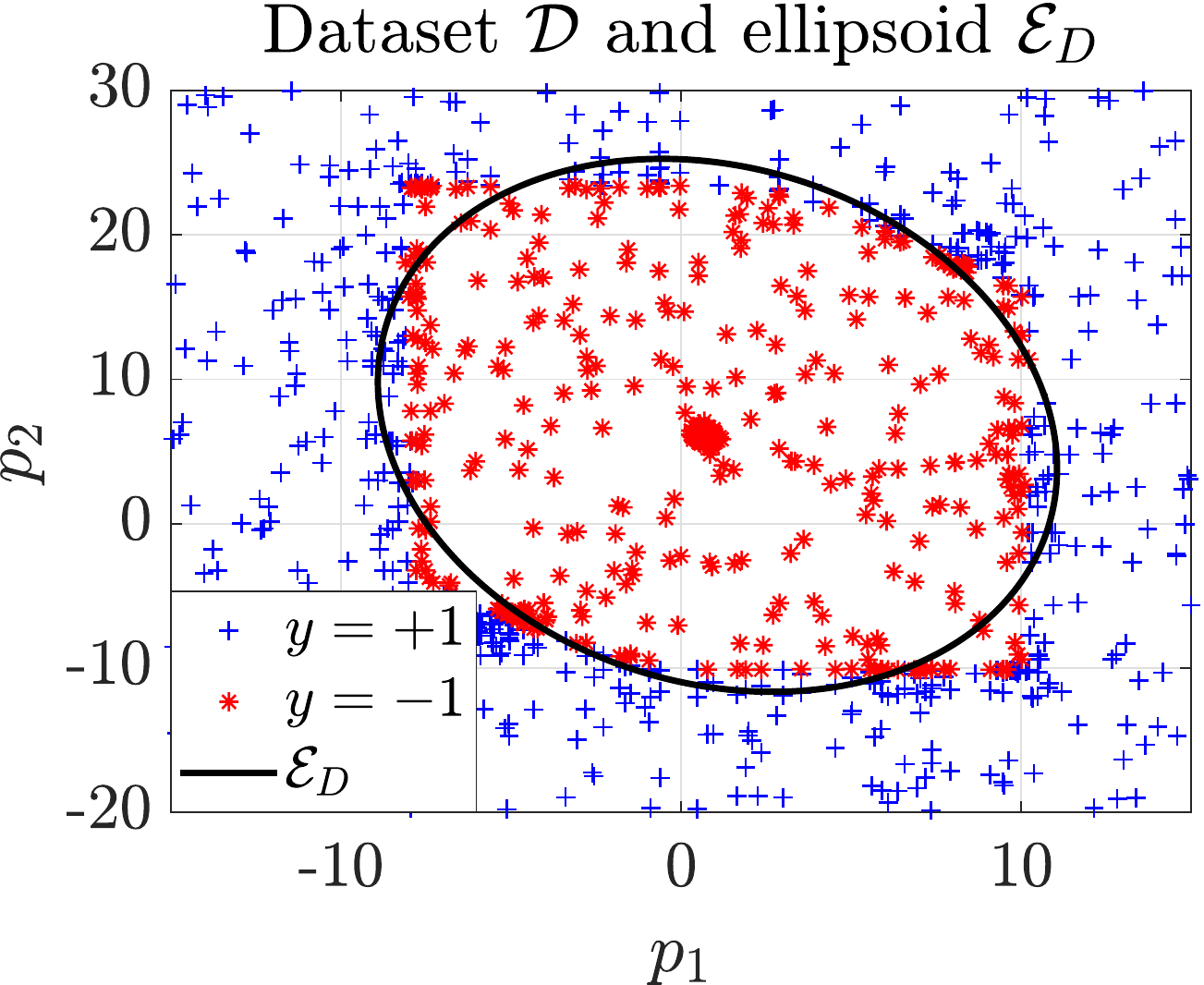}}
	\subfigure{\includegraphics[width=0.24\textwidth]{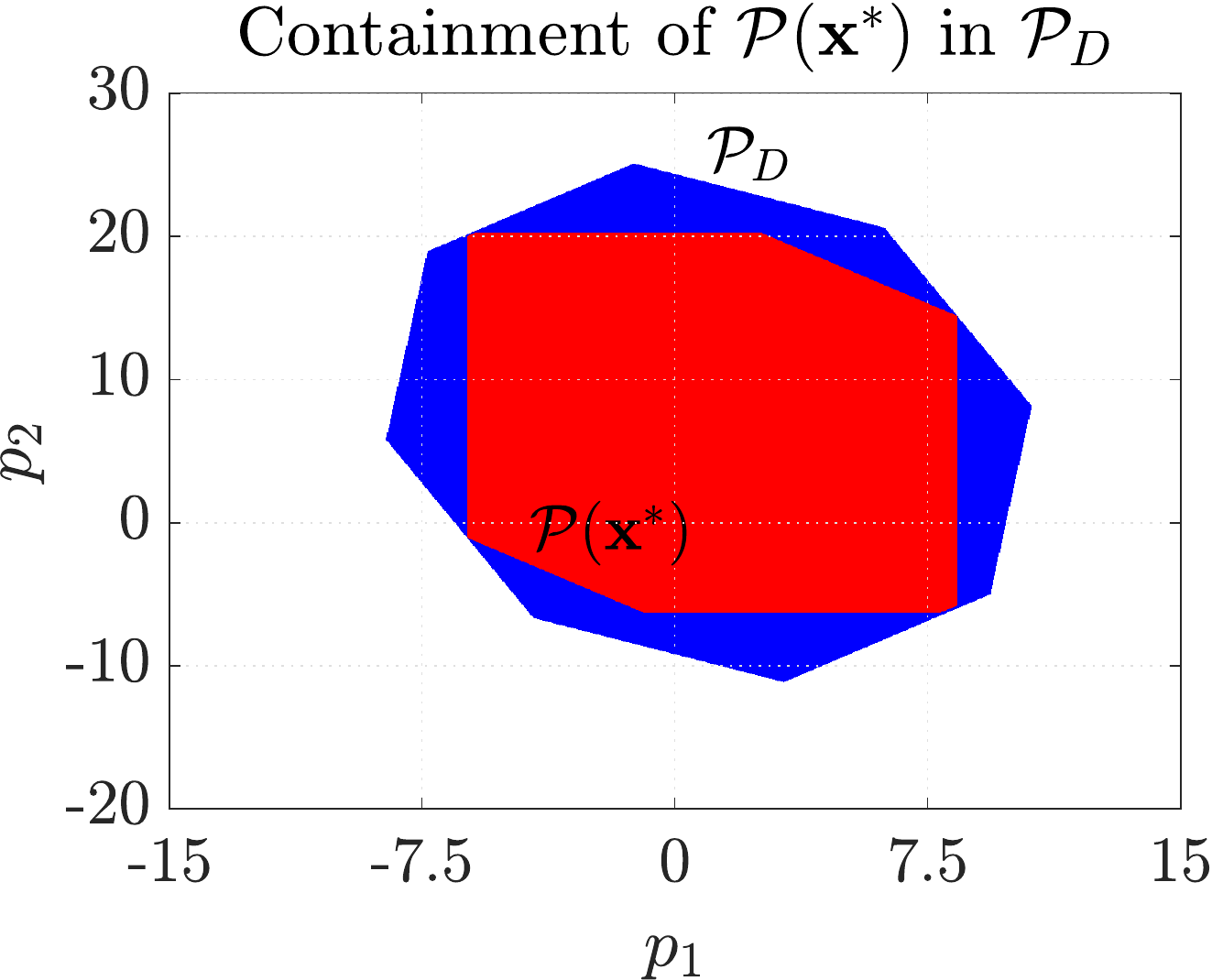}}
	\caption{\emph{Left:} Sampled labeled points of the realistic aggregator and the trained convex quadratic classifier and resulting ellipsoid $\mcE_D$ for $T=2$. \emph{Right:} Inner polytopic approximation $\mcP_D$ of $\mcE_D$, and the final solution $\mcP(\bx^*)$.}
	\label{fig:T=2_real_data}
\end{figure}

\emph{Data generation}: To generate the training dataset $\mcD$, we solved the disaggregation problem in~\eqref{eq:DA} for $T=8$ control periods using the $\ell_1$-norm in the objective to arrive at an MILP. We followed the steps explained in the appendix with $K=25$ and $\kappa=0.2$. 

We first tested the case of $T=2$ to visualize how the proposed OFD performs with realistic data. However, $T=2$ is limiting for control of EVs and for this reason EVs were removed from this particular test. We tried solving the OFD for probability of infeasibility $\epsilon=0.04$, which resulted in a dataset with $N=500$ points ($200$ feasible and $300$ infeasible ones). This dataset and the obtained ellipsoid $\mcE_D$ are shown in Figure~\ref{fig:T=2_real_data} (left). The obtained classifier had a training accuracy of $94.68\%$ and a validation accuracy of $93.76\%$. Figure~\ref{fig:T=2_real_data} (right) shows the inner polytopic approximation $\mcP_D$ of $\mcE_D$, and the final solution $\mcP(\bx^*)$, which demonstrates that the obtained $\mcP(\bx^*)$ closely models the aggregate flexibility.

We continued with the case of $T=8$ and included all 4 device types. We tried solving the OFD for different values of $\epsilon\in\{0,0.04,0.08,0.12\}$, which resulted in four datasets. Each dataset consisted of $N=7,244$ points, equally split between feasible and infeasible ones.

We first wanted to study whether set $\mcF$ is convex by examining dataset $\mcD$ being a snapshot of $\mcF$. If $\mcD^-$ is the subset of $\mcD$ of all feasible points, we evaluated two metrics:
\begin{itemize}
    \item $M_1$: \% of infeasible points in $\mcD$ falling in $\mathrm{conv}(\mcD^-)$.
    \item $M_2$: \% of infeasible points falling in $\mathrm{conv}(\mcD^-)$.
\end{itemize}
Metric $M_1$ can be computed easily by solving an LP for each of the infeasible points in $\mcD$. Nonetheless, this metric can be biased if $\mcD$ has been generated per the process of the appendix. This is because several infeasible points in $\mcD$ have been intentionally selected to lie close to the boundary of $\mcF$, so they have higher chances of falling in $\mathrm{conv}(\mcD^-)$. Metric $M_2$ alleviates this issue but is harder to compute. To approximate $M_2$, we sampled $I=100$ points uniformly at random from $\mathrm{conv}(\mcD^-)$ and checked what percentage of them were infeasible. However labeling each one of these points entails solving \eqref{eq:DA} for $K$ times. Table~\ref{tbl:train} shows the two metrics for the four datasets. Metric $M_2$ is smaller than $M_1$ (in fact zero), as expected. The two metrics indicate that $\mcF$ is described quite accurately by a convex set, which resonates with the analytical findings in~\cite{HSAC21}. Nonetheless, electricity markets may want to surrogate $\mcF$ with polytope $\bG\bp\leq\bx$ for a $\bG$ that is convenient for market clearing processes. We next evaluated our OFD design under this requirement.

\begin{table}[t]
	\renewcommand{\arraystretch}{1.2}
	\caption{Training Results}
	\label{tbl:train}
	\centering
	\begin{tabular}{|c|r|r|r|r|}
		\hline\hline
		&\multicolumn{4}{c|}{Probability of infeasible aggregator schedule $\epsilon$}\\
		\hline
	metric & $0$ &$0.04$ & $0.08$ & $0.12$\\
		\hline\hline
		convexity metric $M_1$&$0.028\%$&$0.145\%$&$0.183\%$&$0.224\%$\\
		\hline
		convexity metric $M_2$&$0.0\%$&$0.0\%$&$0.0\%$&$0.0\%$\\
		\hline
		training accuracy&$97.5\%$&$95.2\%$&$92.5\%$&$91.6\%$\\
		\hline
		validation accuracy&$98.6\%$&$94.6\%$&$92.0\%$&$91.1\%$\\
		\hline
		cond. number of $\bW_2$ &$5.91$ & $5.27$ & $5.05$ & $4.66$\\
		\hline\hline
	\end{tabular}
\end{table}

Following Alg.~\ref{alg:OFD}, we first trained the quadratic classifier by solving \eqref{eq:classifier} using MOSEK and YAMIP in MATLAB. To avoid overfitting, dataset $\mcD$ was partitioned randomly into a training and a validation dataset in a proportion of 80\% to 20\%. Parameter $\lambda$ was tuned to $\lambda=10^{-6}$. Training took consistently less than 1~min. All tests were performed on an Intel Core i7 @ 3.4 GHz (16 GB RAM) computer using MATLAB on a single CPU without any parallelization. Table~\ref{tbl:train} reports the accuracy attained during training and validation by counting the percentage of points correctly labeled by the classifier. Such high accuracy suggests that $\mcF$ can be well approximated by the learned ellipsoid $\mcE_D$. It could also be argued that the two sets have roughly similar volumes. Table~\ref{tbl:train} reports the condition number of $\bW_2$. Having a well-conditioned $\bW_2$ is important in the transformations of Sec.~\ref{subsec:poly}. The volume of $\mcE_D$ can be computed in closed form and is shown in Table~\ref{tbl:Farkas}. 

\begin{table}[t]
	\renewcommand{\arraystretch}{1.2}
	\caption{Results on Optimal Flexibility Design (OFD)}
	\label{tbl:Farkas}
	\centering
	\begin{tabular}{|c|r|r|r|r|}
		\hline\hline
		&\multicolumn{4}{c|}{Probability of infeasible aggregator schedule $\epsilon$}\\
		\hline
	metric & $0$ &$0.04$&$0.08$&$0.12$\\
		\hline\hline
		 \% infeasible $\in\mcP(\bbx)$&$13.64\%$&$19.02\%$&$21.72\%$&$24.39\%$\\
		\hline  $\hat{V}(\bx_i)$&$3.27$&$5.53$ & $5.84$ & $6.88$\\
		\hline
		$\hat{V}(\bx_o)$&$7.23$&$11.86$ & $12.52$ & $21.09$\\
		\hline
		$\hat{V}(\bx_r)$&$0.35$&$0.35$&$0.35$&$2.68$\\
		\hline
		$\hat{V}(\bx_m)$&$53.9$&-&-&-\\
		\hline\hline
	\end{tabular}
\end{table}

Proceeding with the algorithm, we approximated $\mcE_D$ by an inscribed polytope. In Step~2, setting $\delta=0.1$ yielded $23$ extra variables in $\bq$. In Step~3, polytope $\mcP_T^\delta$ was converted to its $\mcP_D$ form using the Fourier-Motzkin algorithm as outlined in Sec.~\ref{subsec:poly}. The obtained matrix $\bE$ of \eqref{eq:PD} had roughly $4.2\cdot 10^{5}$ rows. Steps~2--3 together took no more than $5$~min per dataset. 

For Step~4, we solved~\eqref{eq:xo_minimal} to obtain the related $\bbx$ per dataset, which took less than $1$~min. For Step~5, we solved the LP in~\eqref{eq:maxflex-Farkas2} to obtain the final polytope $\mcP(\bx)$. Given the large size of $\bE$, solving this LP took roughly $4$~hours per dataset. As there is no formula for the volume of a general polytope, the volume of $\mcP(\bx)$ was numerically estimated as detailed next. We first constructed a hyper-rectangle $\mcH=\{\bp:\underline{\bp}\leq \bp\leq\bar{\bp}\}$ containing $\mcP(\bx)$. The $t$-th entry of $\bar{\bp}$ ($\underline{\bp}$) can be computed by maximizing (minimizing) $p_t$ subject to $\bp\in\mcP(\bx)$. The volume of $\mcH$ is $\prod_{t=1}^T(\bar{p}_t-\underline{p}_t)$. We then sampled $10^6$ points from $\mcH$ and found the ratio of these points falling in $\mcP(\bx)$. The estimated volume $\hat{V}(\mcP(\bx))$ was computed by multiplying the ratio times the volume of $\mcH$. The results are presented in Table~\ref{tbl:Farkas}. 

We next studied the effect of ignoring time-coupling by using the reduced matrix $\bG_r^\top=\left[\bI^\top -\bI^\top\right]$. We ran Alg.~\ref{alg:OFD} again and obtained a polytope $\bG_r\bp\leq \bx_r$ whose volume is reported in Table~\ref{tbl:Farkas}. Oversimplifying the aggregator model apparently led to much reduced flexibility. 

To study the effect of uncertainty, we conducted OFD with no chance probability ($\epsilon=0$) assuming externalities in $\bomega$ are set to their mean values. Table~\ref{tbl:Farkas} lists the volume of the computed polytope $\bG\bp\leq \bx_m$, which is significantly larger than that of the polytope computed considering uncertainty. Nevertheless, such flexibility quantification may be misleading as it may not be realizable during operation.

\section{Conclusions}\label{sec:conclusions}
In this work, we have used disaggregation data to capture the time-coupled flexibility of aggregators in presence of random externalities and non-convexity in individual device models. The feasible set of an aggregator has been approximated by the ellipsoid described by a convex quadratic classifier trained over labeled (non)-disaggregatable schedules. As obtaining such labels can be computationally intensive as it involves solving multiple large-scale mixed-integer disaggregation programs, we have put forth an efficient data generation scheme. To arrive from the ellipsoid to the polytopic shape dictated by the market, the ellipsoid is first safely (inner) approximated by a general polytope. Thanks to Farkas' lemma and a volume argument, the general polytope is later approximated by a polytope having the form dictated by the market. Numerical tests corroborate the effectiveness of the novel OFD approach as aggregator's flexibility seems to be well surrogated by a convex set, more detailed polytopes can better inner approximate this feasible set as expected,  while taking into account the time-coupled and uncertain nature of load models seems to be important. 

\appendix
If points $\bp_n$'s are sampled uniformly at random, it is much more likely to sample infeasible than feasible ones. Then to create a balanced dataset $\mcD$, one has to sample a large number of $\bp_n$'s. Albeit sampling points is easy, labeling them can be time consuming. \emph{Labeling a point} entails solving \eqref{eq:DA} for $K$ samples of externalities $\bomega_k$'s to compute the chance statistic in \eqref{eq:DA_test}. We next propose an efficient way for generating a balanced $\mcD$. First, estimate a hyper-rectangle $\mcH$ wherein all feasible schedules could lie. To obtain $\mcH$, one can compute the minimum/maximum values the schedule for each device $\bp^d\in\mcF^d$ can take per control interval. Summing up these values across all devices yields bounds on the aggregator schedule $\bp$. While sampling in $\mcH$ is more efficient than sampling in $\mathbb{R}^T$, obtaining a balanced $\mcD$ can still be challenging as the bounds can be loose. Moreover, infeasible points too far from the boundary of $\mcF$ may not be very informative for classification purposes. In light of these, our goals are: \emph{i)} to sample infeasible points close to the boundary of the feasible set $\mcF$; and \emph{ii)} for each infeasible point we sample, we also identify a feasible point. These goals can be achieved through the following steps:
\renewcommand{\labelenumi}{\emph{D\arabic{enumi})}}
\begin{enumerate}
    \item Sample a point $\bp_1$ uniformly at random within hyper-rectangle $\mcH$. Most likely $\bp_1$ is infeasible, so it is assigned label $y_1=+1$ and pair $(\bp_1,y_1)$ is appended to $\mcD$.
    \item In the process of labeling $\bp_1$, we have also obtained points $\{\hbp(\bomega_k)\}_{k=1}^K$ per \eqref{eq:project}. Among these points, select the one that lies the furthest away from $\bp_1$, that is the one with the largest $g(\hbp(\bomega_k);\bomega_k)$. Call this point $\bp_2$ and label it. If $\bp_2$ is feasible (most likely it is), set $y_2=-1$ and append pair $(\bp_2,y_2)$ to $\mcD$.
    \item Construct point $\bp_3=\kappa\bp_1+(1-\kappa)\bp_2$ for say $\kappa=0.8$, and label it. Most likely, point $\bp_3$ is infeasible but lies closer to the boundary of $\mcF$ than $\bp_1$. This is because the feasible point $\bp_2$ is expected to be close to the boundary of $\mcF$. Append pair $(\bp_3,y_3)$ to $\mcD$. 
    \item Repeat steps \emph{D1)}--\emph{D3)} until $\mcD$ reaches a desirable size.
\end{enumerate}

\begin{figure}
    \centering
    \includegraphics[scale=0.6]{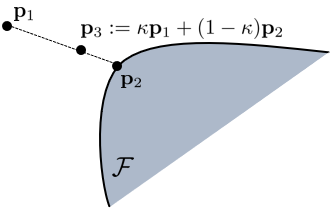}
    \caption{To generate a nearly balanced dataset $\mcD$ in a computationally efficient manner, we draw a triplet of points $(\bp_1,\bp_2,\bp_3)$ at a time. Point $\bp_1$ is drawn uniformly at random and is most likely infeasible. Along its labeling process however, we find point $\bp_2$ that is the approximate projection of $\bp_1$ on feasible set $\mcF$. Point $\bp_3$ is constructed as a convex combination of $(\bp_1,\bp_2)$ lying closer to $\bp_2$. Albeit infeasible, point $\bp_3$ lies close to the boundary of $\mcF$. Sampling such triplets yields a dataset having twice as many infeasible points as feasible ones. Upon taking convex combinations of existing $\bp_2$'s, additional points lying in the interior of $\mcF$ can be constructed.}
    \label{fig:proj}
\end{figure}

Figure~\ref{fig:proj} depicts the triplet $(\bp_1,\bp_2,\bp_3)$ presuming $\bp_1$ lies far away from the boundary of $\mcF$; point $\bp_2$ lies on or close to the boundary; and $\bp_3$ is infeasible but close to the boundary of $\mcF$. The labels assumed for $(\bp_1,\bp_2,\bp_3)$ have been in fact the actual labels obtained for the vast majority of the points sampled during our tests. We next cover the remaining special cases for completeness. If $\bp_1$ is feasible, we can simply add $(\bp_1,y_1)$ to $\mcD$ with $y_1=-1$, and repeat \emph{D1)}. If $\bp_2$ is infeasible, we can replace it with the point $\hbp(\bomega_k)$ obtained during its labeling that lies the furthest away from $\bp_2$. Although there is no analytical guarantee, replacing $\bp_2$ with its projection seems to be reaching a feasible point within one or two iterations. Finally, if $\bp_3$ is feasible, repeat \emph{D3)} and use $\bp_3$ instead of $\bp_2$ in the convex combination to find the new $\bp_3$.

The previous process is expected to generate double the number of infeasible over feasible points. Moreover, all feasible points lie on or close to the boundary of $\mcF$, which can be problematic when training a classifier. Both of these issues can be easily resolved by generating some feasible points that are strictly in the interior of $\mcF$. To this end, we also sample random points as the convex combination of already identified feasible points and label them using the standard process. While such convex combinations are not guaranteed to be feasible, our experiments in Section~\ref{sec:tests} show that they are very likely to be feasible.  

\balance
\bibliographystyle{IEEEtran}
\bibliography{myabrv,power}
\end{document}